\documentclass[prd,twocolumn,nofootinbib]{revtex4}
\usepackage{graphicx, epsfig}
\usepackage{color}
\usepackage{mathrsfs}
\usepackage{bm}
\usepackage{amsmath,amssymb}% for \eqref
\usepackage[caption=false]{subfig}
\usepackage{hyperref}
\usepackage[normalem]{ulem}
\usepackage{mathtools}
\usepackage[x11names]{xcolor}

\definecolor{fashionfuchsia}{rgb}{0.96, 0.0, 0.63}
\colorlet{no_so_fashion_purple}{blue!50!red}

\newcommand{\be}{\begin{equation}}
\newcommand{\ee}{\end{equation}}
\newcommand{\ba}{\begin{eqnarray}}
\newcommand{\ea}{\end{eqnarray}}

\newcommand{\nn}{\nonumber}
\newcommand{\half}{\frac{1}{2}}
\newcommand{\la}{\langle}
\newcommand{\ra}{\rangle}

\newcommand{\bfB}{{\bf B}}

\newcommand{\calE}{{\cal E}}
\newcommand{\bfQ}{{\bf Q}}
\newcommand{\bfW}{{\bf W}}
\newcommand{\bfa}{{\bf a}}

\newcommand{\bfb}{{\bf b}}

\newcommand{\bfp}{{\bf p}}

\newcommand{\bfx}{{\bf x}}
\newcommand{\bfepsilon}{{\bm \epsilon}}
\newcommand{\dbar}{d\hspace*{-0.08em}\bar{}\hspace*{0.1em}}

\def\itie{{\it i.e.}\,}
\def\half{\frac{1}{2}}

\begin{document}
\title{Instability of a uniform electric field in pure non-Abelian Yang-Mills theory}
\author{Carlos Cardona and Tanmay Vachaspati}
\affiliation{
$^*$Physics Department, Arizona State University, Tempe,  Arizona 85287, USA.
}

\begin{abstract}
\noindent
We study the Schwinger process in a uniform non-Abelian electric field using a dynamical
approach in which we evolve an initial quantum state for gluonic excitations.
We evaluate the spectral energy density and number density in the excitations as
functions of time. The total energy density has an ultraviolet divergence which we argue
gets tamed due to asymptotic freedom, leading to $g^4E^4t^4$ growth, where $g$ is 
the coupling and $E$ the electric field strength. 
We also find an infrared divergence in the number density of excitations whose
resolution requires an effect such as confinement.
\end{abstract}

\maketitle

\section{Introduction}
\label{introduction}

The Schwinger effect~\cite{Schwinger:1951nm}, whereby non-perturbative quantum effects 
in a background electric field lead to electron-positron pair production, has received much 
attention (for example, see the reviews ~\cite{doi:10.1119/1.19313,DUNNE_2005, DiPiazza:2011tq,Marklund:2008gj, Dittrich:2000zu}).
Heuristically, the electric field pulls apart the electron-positron pairs
that are fluctuating in and out of the vacuum. For weak electric field strength $E \ll \pi m_e^2/e$, 
where $e$ and $m_e$ are the positron's electric 
charge and mass, the Schwinger process can be thought of as a quantum tunneling event 
and is exponentially suppressed in the dimensionless combination of variables $\pi m_e^2/eE$.
The rate of creation of number density of electrons is,
\be
{\dot n} \propto e^2 E^2 \exp(-\pi m_e^2/eE)
\label{schwinger}
\ee
Schwinger's original computation deals with the probability of vacuum persistence\footnote{For a nice account on the relation between the vacuum persistence probability and rate of pair creation see  \cite{Cohen:2008wz}}
for spin-0 and spin-1/2  matter fields and has 
since been
 generalized to numerous other 
systems besides electromagnetism.
Pair creation has also been studied for 
massless charges,
equivalently for supercritical electric fields ($eE \gg \pi m_e^2$) in 1+1 
dimensions~\cite{Chu:2010xc,Gold:2020qzr} and graphene \cite{Smolyansky:2019dqd,  Fillion-Gourdeau:2015dga,  Gavrilov:2012jk}.
In these cases, the exponential suppression of the original Schwinger effect is not present, 
and other techniques have to be employed as pair creation is no longer a tunneling process 
that is exponentially suppressed. For example, the massless Schwinger model can be
solved completely, including backreaction on the electric field, and results in a $1/\sqrt{t}$ 
decay of the electric field strength~\cite{Chu:2010xc}.

In the present work we are interested in non-Abelian gauge theory in the background
of a homogeneous (color) electric field and the consequent Schwinger pair creation of 
``gluons''\footnote{We
will refer to the model as ``pure QCD'' even though we will consider the simpler 
SU(2) gauge group. The ``pure'' means that we will only consider gauge fields and 
not include any other fields.}.
The process has been investigated before using effective action 
techniques to calculate vacuum persistence amplitudes~\cite{Yildiz:1979vv,Ambjorn:1981qc,Ambjorn:1982nd,Cooper:2005rk,
Nayak:2005yv,Ragsdale:2017wgi, Cooper:2008vy,Matinyan:1976mp,Brown:1979bv,Nair:2010ea}
with the result that there is a constant rate of particle number density production, still 
given by \eqref{schwinger} with $m_e=0$. The result is surprising to us since the Schwinger
process can be viewed as a tunneling process and one might expect that the Wentzel–Kramers–Brillouin (WKB) \cite{Brezin:1970xf,Kim:2000un}
and other approximations used to obtain \eqref{schwinger} 
would break down for massless gauge fields.
For this reason we wish to re-examine the problem using a different approach.

We take a dynamical approach to the problem. (For a kinetic approach to QED, see \cite{Aleksandrov:2020mez,Pervushin:2006vh,Schmidt:1998vi,Hebenstreit:2010vz,Kluger:1998bm,Bialynicki-Birula:1991jwl}.)
At the initial time, we consider a color electric field background
and quantum excitations in their 
non-interacting ground state. We then evolve the system,
just as one would do for small quantum excitations in a time-dependent background. 
We use the method of Bogoliubov transformations~\cite{Bogolyubov:1958,Birrell:1982ix},
re-cast as a ``classical-quantum correspondence'' (CQC) 
whereby quantum evolution is described in terms of the classical evolution of a related
classical system~\cite{Ambjorn:1983ne,Vachaspati:2018llo,Vachaspati:2018hcu,
Bojowald:2020emy}. 
The method is explained in Appendix~\ref{appA}. Our approach simply evolves an
initial state in contrast to other methods that compute (non-interacting) vacuum
persistence amplitudes.

Our results indeed differ from \eqref{schwinger} with $m_e$ set to zero. We find that the
energy density in excitations, ${\cal E}$, grows with time as ${\cal E}\propto g^4E^4 t^4$
where $g$ is the non-Abelian coupling 
constant 
(see also \cite{Akhmedov:2014doa,  Akhmedov:2014hfa}).
We also examine the number density of particles 
in the leading adiabatic approximation~\cite{Dabrowski:2016tsx}
and find that it is not well-defined as there is a zero frequency mode
at all times for which the particle number density diverges. The total number density,
found by integrating over all excitation modes, also diverges. 

In Sec.~\ref{setup} we set up the basic framework, identifying the background field
and the small excitations. Here we also discuss some limitations of our setup.
Sec.~\ref{modes} diagonalizes the Hamiltonian by expanding the quantum fields
in modes. 
We find it convenient to discretize the modes for numerical analysis 
(Sec.~\ref{discretization}). The quantum analysis is reduced to a classical
analysis in Sec.~\ref{cqc}, following which we numerically evaluate the 
spectral energy density and the total energy in quantum excitations as a function 
of time in Sec.~\ref{energydensity}.
The number density of quantum excitations is discussed in Sec.~\ref{numberdensity}
and shown to diverge for all times due to the presence of a zero frequency mode.  Sec.  ~\ref{adiabatic} briefly considers the case of adiabatically turning on the external field.
We conclude in Sec.~\ref{conclusions}.

\section{Setup}
\label{setup}

We will consider a pure $SU(2)$ gauge theory with Lagrangian density,
\be
L = -\frac{1}{4} (W^a_{\mu\nu})^2, \ \ a=1,2,3
\label{su2lag}
\ee
with the field strength defined in the usual way
\be
W^a_{\mu\nu} = \partial_\mu W^a_\nu - \partial_\nu W^a_\mu
+ g \epsilon^{abc} W^b_\mu W^c_\nu
\ee
We have to make certain approximations to proceed with our analysis.
Our main approximation is that we expand the fields about a fixed background electric 
field to quadratic order in the Lagrangian and ignore higher order interactions. 

Let us write
\be
W^a_\mu = A^a_\mu + Q^a_\mu
\label{perts}
\ee
where $A^a_\mu$ is a classical background and $Q^a_\mu$ denotes
quantum excitations on top of the classical background. We will work in
temporal gauge, so $W^a_0=0$, and take
\be
A^a_\mu = - A(t) \, \delta^{a3} \delta_{\mu 3}
\label{uniformbgnd}
\ee
Then there is an externally imposed classical electric field but 
no magnetic field,
\be
%E^a_i = - A^a_{0i} = {\dot A}(t) \, \delta^{a3} \delta_{i3}, \ \  B^a_i = 0.
E^a_i = - \partial_t A^a_i = {\dot A}(t) \, \delta^{a3} \delta_{i3}, \ \  B^a_i = 0.
\ee

We will evolve the quantum variables, $Q^a_i$, assuming that they are
in their non-interacting ground state at $t=0$. 
We insert \eqref{perts} in \eqref{su2lag}, then use the background \eqref{uniformbgnd}, 
and expand to quadratic order in the $Q^a_\mu$ to obtain,
\ba
L &=& 
\frac{1}{2} (\dot Q^1_i)^2
-\frac{1}{4} (\partial_i Q^1_j - \partial_j Q^1_i - g A (Q^2_i \delta^3_j - \delta^3_i Q^2_j))^2 
\nn \\
&+& \frac{1}{2} (\dot Q^2_i)^2
-\frac{1}{4} (\partial_i Q^2_j - \partial_j Q^2_i + g A (Q^1_i \delta^3_j - \delta^3_i Q^1_j))^2 \nn \\
&+& \frac{1}{2} ({\dot Q}^3_3 - {\dot A}) ^2 - \frac{1}{4} (\partial_i Q^3_j - \partial_j Q^3_i)^2
+ {\cal O}\left ( (Q^a_i)^3 \right )
\label{expandedL}
\ea
The classical electric field is externally imposed, \itie there are external sources that
produce and maintain the electric field $E^3_3$ which is assumed to be constant.
Therefore, we take\footnote{We will also consider an adiabatically turned on and off
electric field in Sec.~\ref{adiabatic}.}
\be
A(t) = E t, \ \ {\dot A}= E
\ee
Then the variables $Q^3_i$ decouple from the other quantum variables. We can calculate the 
rate of particle production in a fixed external field by considering the truncated Lagrangian,
\be
L' = 
\frac{1}{2} (\dot Q^1_i)^2 + \frac{1}{2} (\dot Q^2_i)^2 - \frac{1}{4} (Q^1_{ij})^2 - \frac{1}{4} (Q^2_{ij})^2
\label{Ltruncated}
\ee
where,
\ba
Q^1_{ij} &\equiv& \partial_i Q^1_j - \partial_j Q^1_i - g E t (Q^2_i \delta^3_j - \delta^3_i Q^2_j) \nn \\
Q^2_{ij} &\equiv& \partial_i Q^2_j - \partial_j Q^2_i + g E t (Q^1_i \delta^3_j - \delta^3_i Q^1_j)
\ea
At this level of approximation, the $Q^1_i$ and $Q^2_i$ fluctuations do not backreact on the
background electric field. The backreaction will only appear due to the cubic and higher order 
terms in the $Q^a_i$ in \eqref{expandedL}.

We will expand the excitations in momentum modes in the next section. There are quantum
issues at both ends of the spectrum. For modes with low energy, the coupling constant is strong 
and confinement should play a role. The lowest energy excitations will be massive glueballs, not
massless gluons\footnote{Indeed the assumed background uniform electric field itself ignores
confinement. The spirit of the present work is that we work as if there is no confinement and
study the consequences.}.
Modes with very high energy are in the regime of asymptotic freedom as
the coupling constant becomes small. Inclusion of these effects in our calculations is
beyond our reach and we shall proceed based on \eqref{Ltruncated} as if it is the full
story and see if there are any inconsistencies.
Indeed we will encounter two inconsistencies in this approach.
In Sec.~\ref{numericalenergy} we will encounter an ultraviolet divergence that we 
argue will be resolved by properly accounting for asymptotic freedom.
We will also encounter a divergence in the number density
of excitations at low energy at all times whose interpretation will change radically 
once we take confinement into account.

\section{Expansion in modes}
\label{modes}

In this section we expand the fields in modes and diagonalize the Lagrangian. The
calculations are straightforward if tedious;
the end result for the diagonalized Lagrangian is given in \eqref{L1} and \eqref{L2}.

The gauge fields are expanded in the physical transverse modes as,
\be
Q^1_\mu = \int \frac{\dbar ^3 p}{\sqrt{2E_p}} \sum_{r=0}^3 \left [ 
a^r_\bfp \epsilon_{\bfp,\mu}^r e^{i\bfp\cdot\bfx} + h.c. \right ]
\ee
where $\dbar^3p \equiv d^3p/(2\pi)^3$, $E_p=|\bfp |$, and  the reality of 
$Q^1$ implies $a^{r*}_\bfp=a^r_{-\bfp}$. Similarly,  
\be
Q^2_\mu = \int \frac{\dbar ^3 p}{\sqrt{2E_p}} \sum_{r=0}^3 \left [ 
b^r_\bfp \epsilon_{\bfp,\mu}^r e^{i\bfp\cdot\bfx} + h.c. \right ]
\ee
with $b^{r*}_\bfp=b^r_{-\bfp}$.

The magnetic fields are
\ba
(\bfB^1)_i &=& \frac{1}{2} \epsilon_{ijk} Q^1_{jk} = \nabla \times \bfQ^1 - gE t \, \bfQ^2 \times {\hat z} \\
(\bfB^2)_i &=& \frac{1}{2} \epsilon_{ijk} Q^2_{jk} =\nabla \times \bfQ^2 + gE t \, \bfQ^1 \times {\hat z}
\ea
Therefore
\ba
\bfB^1 &=& 
\int \frac{\dbar ^3 p}{\sqrt{2E_p}} \sum_{r=0}^3 \left [ 
( a^r_\bfp i \bfp  + gEt \, b^r_\bfp {\hat z} ) \times \bfepsilon^r_\bfp e^{i\bfp\cdot\bfx} + h.c. \right ]
\nn \\
&\equiv&
\int \frac{\dbar ^3 p}{\sqrt{2E_p}} \sum_{r=0}^3 \left [ 
\bfa^r_\bfp e^{i\bfp\cdot\bfx} + h.c. \right ]
\ea
where $\bfa^r_\bfp \equiv ( a^r_\bfp i \bfp  + gEt \, b^r_\bfp {\hat z} ) \times \bfepsilon^r_\bfp$,  and
\ba
\bfB^2 &=& \int \frac{\dbar ^3 p}{\sqrt{2E_p}} \sum_{r=0}^3 \left [ 
( b^r_\bfp i \bfp  - gEt \, a^r_\bfp {\hat z} ) \times \bfepsilon^r_\bfp e^{i\bfp\cdot\bfx} + h.c. \right ]
\nn \\
&\equiv&
\int \frac{\dbar ^3 p}{\sqrt{2E_p}} \sum_{r=0}^3 \left [ 
\bfb^r_\bfp e^{i\bfp\cdot\bfx} + h.c. \right ]
\ea
where $\bfb^r_\bfp \equiv ( b^r_\bfp i \bfp  - gEt \, a^r_\bfp {\hat z} ) \times \bfepsilon^r_\bfp$.

These expressions give,
\ba
\calE_{B1} &\equiv& \half \int d^3x (\bfB^1)^2 \nn \\
&=& \int \frac{\dbar ^3 p}{2E_p} 
\sum_{r,s=0}^3 \left [ \bfa^r_\bfp \cdot \bfa^{s\dag}_\bfp + h.c. \right ]
\ea
Note that there are two polarizations and $\{ {\hat \epsilon}^1_\bfp,{\hat \epsilon}^2_\bfp, {\hat p} \}$ form
a right-handed orthonormal basis. For example,
\ba
{\hat \epsilon}^1_\bfp &=& (-\cos\theta \cos\phi, -\cos\theta \sin\phi, \sin\theta ) 
\label{hateps1} \\
{\hat \epsilon}^2_\bfp &=& (-\sin\phi , \cos\phi, 0 ) 
\label{hateps2} \\
{\hat p} &=& (\sin\theta \cos\phi, \sin\theta \sin\phi, \cos\theta ) 
\label{hatp}
\ea

We check
\be
(\bfp \times {\hat \epsilon}^r_\bfp) \cdot (\bfp \times {\hat \epsilon}^s_\bfp) = p^2 \delta^{rs}
\ee
\be
({\hat z} \times {\hat \epsilon}^r_\bfp) \cdot (\bfp \times {\hat \epsilon}^s_\bfp) = p_z \delta^{rs}
= p\cos\theta \delta^{rs}
\ee
\be
({\hat z} \times {\hat \epsilon}^r_\bfp) \cdot ({\hat z} \times {\hat \epsilon}^s_\bfp) = 
\delta^{rs} - ({\hat \epsilon}^r_\bfp \cdot {\hat z}) \, ({\hat \epsilon}^s_\bfp \cdot {\hat z})
\ee
and with the choice of vectors in \eqref{hateps1}-\eqref{hatp},
\be
 ({\hat \epsilon}^r_\bfp \cdot {\hat z}) \, ({\hat \epsilon}^s_\bfp \cdot {\hat z}) = 
\sin^2\theta \, \delta^{r1}\delta^{s1}
\ee
Therefore
\ba
\calE_{B1} &=& \int \frac{\dbar ^3 p}{E_p} 
\sum_{r=1}^2 \biggl [ 
p^2 |a^r_\bfp |^2 + g^2E^2 t^2 |b^r_\bfp |^2 (1-\sin^2\theta \delta_{r1} ) \nn \\
&& \hskip 2 cm
- i gE t p_z (a^{r\dag}_\bfp b^r_\bfp - a^r_\bfp b^{r\dag}_\bfp )
\biggr ]
\ea
Similarly
\ba
\calE_{B2} &=& \int \frac{\dbar ^3 p}{E_p} 
\sum_{r=1}^2 \biggl [ 
p^2 |b^r_\bfp |^2 + g^2E^2 t^2 |a^r_\bfp |^2 (1-\sin^2\theta \delta_{r1} ) \nn \\
&& \hskip 2 cm
- i gE t p_z (a^{r\dag}_\bfp b^r_\bfp - a^r_\bfp b^{r\dag}_\bfp )
\biggr ]
\ea

\ba
\calE_{B1+2} &=& \calE_{B1}+\calE_{B2} \nn \\
&=& \int \frac{\dbar ^3 p}{E_p} \biggl [ 
(p^2 + g^2E^2 t^2 \cos^2\theta) ( |a^1_\bfp |^2 + |b^1_\bfp |^2 ) \nn \\
&& \hskip 1 cm
- i 2 gE t p \cos\theta (a^{1\dag}_\bfp b^1_\bfp - a^1_\bfp b^{1\dag}_\bfp ) \nn \\
&& \hskip 1 cm
+ (p^2 + g^2E^2t^2) ( |a^2_\bfp |^2 + |b^2_\bfp |^2 ) \nn \\
&& \hskip 1 cm
- i 2 gE t p \cos\theta (a^{2\dag}_\bfp b^2_\bfp - a^2_\bfp b^{2\dag}_\bfp )
\biggr ]
\ea

Next let
\be
a^r_\bfp = \alpha^r_\bfp +i\beta^r_\bfp, \ \ 
b^r_\bfp = \gamma^r_\bfp +i\delta^r_\bfp
\ee
where $\alpha^r_\bfp$, $\beta^r_\bfp$, $\gamma^r_\bfp$ and $\delta^r_\bfp$ are real. 
The reality conditions, $a^{r*}_\bfp=a^r_{-\bfp}$ and $b^{r*}_\bfp=b^r_{-\bfp}$, imply
\be
\alpha^r_\bfp = \alpha^r_{-\bfp}, \ \  \beta^r_\bfp = -\beta^r_{-\bfp}, \ \ 
\gamma^r_\bfp = \gamma^r_{-\bfp}, \ \ \delta^r_\bfp = - \delta^r_{-\bfp}
\ee
For convenience, define 
\be
P=gE t, \ \  P_z=P\cos\theta
\label{PPz}
\ee
Then
\ba
\calE_{B1+2}  &=& \int \frac{\dbar ^3 p}{E_p} \biggl [
(p^2 + P_z^2) \{ (\alpha^1_\bfp)^2 + (\delta^1_\bfp)^2 \} + 4 p P_z \alpha^1_\bfp \delta^1_\bfp \nn \\
&& \hskip 0.5 cm
+ (p^2 + P_z^2) \{ (\beta^1_\bfp)^2 + (\gamma^1_\bfp)^2 \} - 4 p P_z \beta^1_\bfp\gamma^1_\bfp \nn \\
&& \hskip 0.5 cm
+ (p^2 + P^2) \{ (\alpha^2_\bfp)^2 + (\delta^2_\bfp)^2 \} + 4 p P_z \alpha^2_\bfp\delta^2_\bfp \nn \\
&& \hskip 0.5 cm
+ (p^2 + P^2) \{ (\beta^2_\bfp)^2 + (\gamma^2_\bfp)^2 \} - 4 p P_z \beta^2_\bfp\gamma^2_\bfp
\biggr ] \nn
\ea
The energy has separated into four disjoint sectors: ($\alpha^1_\bfp,\delta^1_\bfp$),
($\beta^1_\bfp,\gamma^1_\bfp$), ($\alpha^2_\bfp, \delta^2_\bfp$) and ($\alpha^2_\bfp,\delta^2_\bfp$).
The ($\alpha^1_\bfp,\delta^1_\bfp$) sector is equivalent to the ($\beta^1_\bfp,\gamma^1_\bfp$) sector
under $P_z \to -P_z$ which is the same as $g \to -g$.  Similarly,  for the ($\alpha^2_\bfp, \delta^2_\bfp$)
and ($\beta^2_\bfp,\gamma^2_\bfp$) sectors.
The ($\alpha^1_\bfp,\delta^1_\bfp$) and ($\alpha^2_\bfp,\delta^2_\bfp$) sectors look very similar
but they differ in their first terms: the former has $P_z$ while the latter has $P$.

The energy in the electric field is
\ba
\calE_{E1+2} &=& \half \int d^3x \left [ (\partial_t \bfW^1)^2 + (\partial_t \bfW^2)^2 \right ] \nn \\
&=& \int \frac{\dbar^3p}{E_p} \sum_{r=1}^2 \left [ ( {\dot \alpha}^r_\bfp )^2 + ({\dot \beta}^r_\bfp)^2
+ ( {\dot \gamma }^r_\bfp )^2 + ({\dot \delta }^r_\bfp )^2 \right ] \nn
\ea
and we only need to solve for the dynamics of the two quantum systems,
\ba
L_1 &=& \int \frac{\dbar^3p}{E_p} \biggl [ ( {\dot \alpha}^1_\bfp )^2 + ({\dot \delta }^1_\bfp )^2 
- (p^2 + P_z^2) \{ (\alpha^1_\bfp)^2 + (\delta^1_\bfp)^2 \}
\nn \\ && \hskip 4 cm
- 4 p P_z \alpha^1_\bfp \delta^1_\bfp \biggr ]
\ea
\ba
L_2 &=& \int \frac{\dbar^3p}{E_p} \biggl [ ( {\dot \alpha}^2_\bfp )^2 + ({\dot \delta }^2_\bfp )^2 
- (p^2 + P^2) \{ (\alpha^2_\bfp)^2 + (\delta^2_\bfp)^2 \}
\nn \\ && \hskip 4 cm
- 4 p P_z \alpha^2_\bfp \delta^2_\bfp \biggr ]
\ea
As noted above, the Lagrangians for ($\beta^r_\bfp,\gamma^r_\bfp$) are related to
$L_1$ and $L_2$ by $g \to -g$. Also recall that there is time-dependence in these
Lagrangians because $P_z$ and $P$ grow in proportion to $t$ as defined in \eqref{PPz}.

The Lagrangians can be diagonalized by using linear combinations,
\be
\phi_{\bfp,\pm} = \frac{\alpha^1_\bfp \pm \delta^1_\bfp}{\sqrt{2}}, \ \ 
\psi_{\bfp,\pm} = \frac{\alpha^2_\bfp \pm \delta^2_\bfp}{\sqrt{2}}
\label{phipsi}
\ee
Then
\ba
L_1 &=& \int \frac{\dbar^3p}{E_p} \biggl [ 
{\dot \phi}_{\bfp,+}^2 - (p+P_z)^2 \phi_{\bfp,+}^2  \nn \\
&& \hskip 1.5 cm
+ {\dot \phi}_{\bfp,-}^2 - (p-P_z)^2 \phi_{\bfp,-}^2 \biggr ]
\label{L1}
\ea
\ba
L_2 &=& \int \frac{\dbar^3p}{E_p} \biggl [ 
{\dot \psi}_{\bfp,+}^2 
- \{ (p_z+P)^2+p_\perp^2 \} \psi_{\bfp,+}^2   \nn \\
&& \hskip 1 cm
+ {\dot \psi}_{\bfp,-}^2 - \{ (p_z-P)^2+p_\perp^2 \} \psi_{\bfp,-}^2 \biggr ]
\label{L2}
\ea
where $P=gE t$, $p_\perp= p \sin\theta$ and $p_z=p\cos\theta$.

Similarly,  we can easily write down $L_3$ and $L_4$ for the 
$(\beta^r_\bfp,\gamma^r_\bfp)$ sector since these are related
to $L_1$ and $L_2$ by $g \to -g$. The full Lagrangian is given by
the sum of $L_1,\ldots, L_4$.

\section{Discretization}
\label{discretization}

We now discretize the integrations in \eqref{L1} and \eqref{L2}. For example,
\ba
L_1 &=& \sum_{\bfp} \frac{(\Delta p)^3}{E_p} \biggl [ 
{\dot \phi}_{\bfp,+}^2 - (p+P_z)^2 \phi_{\bfp,+}^2  \nn \\
&& \hskip 0.25 cm
+ {\dot \phi}_{\bfp,-}^2 - (p-P_z)^2 \phi_{\bfp,-}^2 \biggr ] =\sum_{\bfp} L_{1\bfp}
\ea
and similarly for $L_2$. The volume element in momentum space is
\be
(\Delta p )^3 = \left ( \frac{2\pi}{L} \right )^3 =  \frac{(2\pi )^3}{V}
\ee
where $L$ is the size of the (compactified) spatial domain and $V$ is its volume.
The Hamiltonian for each mode can now be written as that for a simple harmonic
oscillator,
\ba
H_{1\bfp} &=& 
\frac{\pi_{\bfp +}^2}{2 m_\bfp} + \frac{m_\bfp}{2} (p+P_z)^2 |\phi_{\bfp +}|^2 \nn \\
&& 
+\frac{\pi_{\bfp -}^2}{2 m_\bfp} + \frac{m_\bfp}{2} (p-P_z)^2 |\phi_{\bfp -}|^2
\label{H1bfp}
\ea
\ba
H_{2\bfp} &=& 
\frac{\Pi_{\bfp +}^2}{2 m_\bfp} + \frac{m_\bfp}{2} 
 \{ (p_z+P)^2+p_\perp^2 \}  |\psi_{\bfp +}|^2 \nn \\
&& 
+\frac{\Pi_{\bfp -}^2}{2 m_\bfp} + \frac{m_\bfp}{2} 
 \{ (p_z-P)^2+p_\perp^2 \} |\psi_{\bfp -}|^2
 \label{H2bfp}
\ea
where $\pi_{\bfp \pm}$ and $\Pi_{\bfp \pm}$ are conjugate momenta to $\phi_{\bfp \pm}$
and $\psi_{\bfp \pm}$, and
\be
m_{\bfp} = 2 \frac{(\Delta p)^3}{E_p} = \frac{2}{p} \left ( \frac{2\pi}{L} \right )^3
\label{mbfp}
\ee
Note that $m_\bfp$ has dimensions of mass squared (not mass).

As noted at the end of Sec.~\ref{modes}, we will also have $H_{3\bfp}$ and $H_{4\bfp}$
corresponding to the $(\beta^r_\bfp,\gamma^r_\bfp)$ sector.

The Hamiltonians in \eqref{H1bfp} and \eqref{H2bfp} are those of simple
harmonic oscillators with time dependent frequencies,
\be
\omega_{\phi,\bfp,\pm} = +\sqrt{(p\pm P_z )^2}
\label{omphi}
\ee
\be
\omega_{\psi,\bfp,\pm} = +\sqrt{(p_z \pm P)^2+p_\perp^2}
\label{ompsi}
\ee
The $+$ signs in the pre-factor are to emphasize that we are taking the positive 
square root. The structure of the frequencies is easy to understand because the 
$p_z\pm P$ follows from the covariant derivative acting on the excitations.
The different forms of $\omega_{\phi,\bfp,\pm}$ and $\omega_{\psi,\bfp,\pm}$ 
arise since the variables $\phi_{\bfp,\pm}$ are associated with ${\hat \epsilon}_\bfp^{r=1}$, 
while $\psi_{\bfp,\pm}$ are associated with ${\hat \epsilon}_\bfp^{r=2}$.

An important point for us is that there are certain modes for which the frequency
vanishes.  For example, $\omega_{\phi,\bfp,-} = |p - P_z |$ and,  at any time, 
the frequency vanishes for $p=P_z = gEt\cos\theta$.
(Similarly for $\omega_{\psi,\bfp,-}$.)
The frequencies $\omega_{\phi,\bfp,+}$ and $\omega_{\psi,\bfp,+}$
do not vanish for $t > 0$. We will return to this point in our discussion of the
particle number density in Sec.~\ref{numberdensity}.

\section{Classical-Quantum Correspondence}
\label{cqc}

To obtain particle production of the fields $\phi_{\bfp,\pm}$, $\psi_{\bfp,\pm}$ we will use
the CQC. (See Appendix~\ref{appA} for a summary of the CQC.)
Then the variables $\phi_{\bfp,\pm}$
and $\psi_{\bfp,\pm}$ are complexified and we solve the classical equations of motion
\ba
{\ddot \phi}_{\bfp,\pm} + (P_z \pm p )^2 \phi_{\bfp,\pm} &=& 0 
\label{phieq} \\
{\ddot \psi}_{\bfp,\pm} + 
 \{ (P \pm p_z)^2+p_\perp^2 \} \psi_{\bfp,\pm} &=& 0 
 \label{psieq}
\ea
 with initial conditions,
 \ba
&& \phi_{\bfp,\pm} (t=0) = - \frac{i}{\sqrt{2m_{\bfp} p}} =\psi_{\bfp,\pm} (t=0) \\
&& {\dot \phi}_{\bfp,\pm} (t=0) = \sqrt{\frac{p}{2m_{\bfp}}} ={\dot \psi}_{\bfp,\pm} (t=0)
 \ea
 (Note that the simple harmonic oscillator frequencies at $t=0$ are simply $p$ because
 $P(t=0)=0$.)
Using \eqref{mbfp} we write
 \ba
&& \phi_{\bfp,\pm} (0) = - \frac{i}{2} \left ( \frac{L}{2\pi} \right )^{3/2} =\psi_{\bfp,\pm} (0) \\
&& {\dot \phi}_{\bfp,\pm} (0) = \frac{p}{2} \left ( \frac{L}{2\pi} \right )^{3/2} ={\dot \psi}_{\bfp,\pm} (0)
\label{dotic}
\ea
 
 To calculate the energy density in excitations, we simply need to evaluate the 
 classical energy in the complexified $\phi_{\bfp,\pm}$ and $\psi_{\bfp,\pm}$ as
 we describe next.

\section{Energy density production}
\label{energydensity}
 
 The energy in the complexified variables $\phi_{\bfp,\pm}$ $\psi_{\bfp,\pm}$
 follows from \eqref{L1} and \eqref{L2},
 \ba
\calE_1 &=& \int \frac{\dbar^3p}{E_p}\biggl [ 
| {\dot \phi}_{\bfp,+} |^2 + (p+P_z)^2 | \phi_{\bfp,+}| ^2 \nn \\ 
&& \hskip 1.5 cm
+ | {\dot \phi}_{\bfp,-}|^2 + (p-P_z)^2 |\phi_{\bfp,-}|^2 \biggr ]
\label{calE1}
\ea 
\ba
\calE_2 &=& \int \frac{\dbar^3p}{E_p} \biggl [ 
| {\dot \psi}_{\bfp,+} |^2 
+ \{ (p_z+P)^2+p_\perp^2 \} | \psi_{\bfp,+} |^2   \nn \\
&& \hskip 0.5 cm
+ | {\dot \psi}_{\bfp,-} |^2 + \{ (p_z-P)^2+p_\perp^2 \} |\psi_{\bfp,-} |^2 \biggr ]
\label{calE2}
\ea
The energies in the $(\beta^r_\bfp,\gamma^r_\bfp)$ sector give identical
expressions and we will include these in the end in the total energy by
multiplying by a factor of two.
 
The energy expressions in \eqref{calE1} and \eqref{calE2} include the
ground state energy --  the $\omega /2$ of the simple harmonic
oscillator -- whereas we are interested in the energy of the excitations only.
As described in Appendix~\ref{appA}, the ground state energy can be
discarded by writing the energies as, 
\ba
:\calE_1: &=& \int \frac{\dbar^3p}{E_p}\biggl [ 
\left | {\dot \phi}_{\bfp,+}  - i \omega_{\phi,\bfp,+}  \, \phi_{\bfp,+} \right | ^2 \nn \\ 
&& \hskip 1.5 cm
+ \left | {\dot \phi}_{\bfp,-} - i \omega_{\phi,\bfp,-} \, \phi_{\bfp,-} \right |^2 \biggr ]
\label{calE1NO}
\ea 
\ba
:\calE_2: &=& \int \frac{\dbar^3p}{E_p} \biggl [ 
\left | {\dot \psi}_{\bfp,+} -i \omega_{\psi,\bfp,+} \, \psi_{\bfp,+}\right  |^2   \nn \\
&& \hskip 1.5 cm
+ \left | {\dot \psi}_{\bfp,-} - i \omega_{\psi,\bfp,-} \, \psi_{\bfp,-} \right |^2 \biggr ]
\label{calE2NO}
\ea
The total energy in the excitations is
\ba
:\calE: &=& 2 (\, :\calE_1: + :\calE_2: \, ) \nn \\
&=& 8\pi \int_0^\infty dp\, p \int_0^1 du\,   \nn \\
&& \hskip -2. cm
\times
\biggl [ \left | {\dot \phi}_{\bfp,+}  - i \omega_{\phi,\bfp,+}  \, \phi_{\bfp,+} \right | ^2 
+ \left | {\dot \phi}_{\bfp,-} - i \omega_{\phi,\bfp,-} \, \phi_{\bfp,-} \right |^2  \nn \\
&& \hskip -2. cm
+ \left | {\dot \psi}_{\bfp,+} -i \omega_{\psi,\bfp,+} \, \psi_{\bfp,+}\right  |^2   
+ \left | {\dot \psi}_{\bfp,-} - i \omega_{\psi,\bfp,-} \, \psi_{\bfp,-} \right |^2 \biggr ]
\label{totalENO}
\ea
where the factor of two in the first line 
accounts for the excitations in the $(\beta^r_\bfp,\gamma^r_\bfp)$ sector,
$u\equiv \cos\theta$, and we have used the symmetry under $u \to -u$ to
restrict $u$ to the interval $[0,1]$.
We remark that the expressions occurring in the integrands of \eqref{calE1NO} 
and \eqref{calE2NO}, such as ${\dot \phi}_{\bfp,+}  - i \omega_{\phi,\bfp,+}  \, \phi_{\bfp,+}$,
are the usual Bogolyubov $\beta$ coefficients up to a factor of $1/\sqrt{\omega}$, where 
$\omega$ stands for the frequency. 

Now all that is required is to solve the equations of motion in \eqref{phieq}
and \eqref{psieq}, insert the solutions in the energy expressions above, and
perform the integrations. The first step is formally accomplished since the solutions
to the equations of motion can be written in terms of parabolic cylindrical functions. 
However, we have found it more practical to solve the differential equations
numerically followed by numerical integration.
%\footnote{Part of the numerical work 
%has been performed in Python and is available at
%\url{https://github.com/cargicar/Vacuum_inestability_Non_Abelian_YM/blob/96625bc7e89863901ba59439d2427def826cae0b/NASchwingerEffect4_carlos_kernel.ipynb}
%}.

\subsection{Numerical evaluation of the energy}
\label{numericalenergy}

We have numerically solved the differential equations in \eqref{phieq} and \eqref{psieq} for
$u\in [0,1]$ and $p \in [0,p_c]$ where $p_c \gg gE t$ is a momentum cutoff.
As we solve the differential equations, we also numerically evaluate the energy in 
\eqref{totalENO}. In the numerical computations we choose units so that $gE=1$.

\begin{figure}
      \includegraphics[width=0.45\textwidth,angle=0]{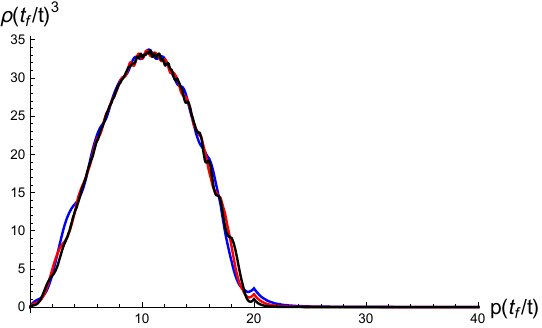}
\caption{
$\rho(x,t) (t_f/t)^3$ vs. $x=p(t_f/t)$ for $t_f=20$ and $t=t_f/2,2t_f/3$, and $t_f$. There are
three curves in the plot but they all overlap. The peak in $\rho(p,t)$ is located at
$p \approx gE t/2$.
}
\label{rhovsp}
\end{figure}

We define the spectral energy density, $\rho (p,t)$, via
\be
\frac{:\calE:}{L^3}  = \int_0^\infty dp\, \rho(p,t)
\ee
or explicitly,
\ba
\rho (p) &=& \frac{8\pi p}{L^3} \int_0^1 du\,  
\sum_{s=\pm} \biggl [ 
 \left | {\dot \phi}_{\bfp,s}  - i \omega_{\phi,\bfp,s}  \, \phi_{\bfp,s} \right | ^2 \nn \\
&& \hskip 2 cm
 + \left | {\dot \psi}_{\bfp,s} -i \omega_{\psi,\bfp,s} \, \psi_{\bfp,s}\right  |^2 \biggr ]
 \label{rhoexplicit}
 \ea

In Fig.~\ref{rhovsp} we plot the spectral energy density rescaled by $t^{-3}$, 
{\it i.e.} $t^{-3} \rho (p,t)$, as a function of $p/t$ at three different times. It is
clear that the peak of the spectrum $\rho$ grows as $t^3$ and the width
grows as $t$, implying that the total energy in excitations grows as $t^4$.

The linear growth of the width of $\rho(p,t)$ follows from the form of the oscillation 
frequencies in \eqref{omphi} and \eqref{ompsi}. For very large values of $p$, the
time dependence of the frequencies can be ignored. Then the solutions
for $\phi_{\bfp,\pm}$ and $\psi_{\bfp,\pm}$ are simply trigonometric
functions for which there is no contribution to the energy. Hence there
is no particle production for $p \gg gE t$; there is only particle production
for $p \lesssim gEt$ and so the width in $p$ contributing
to particle production grows linearly in $t$. To understand the growth of the
peak of $\rho(p,t)$ that goes as $t^3$, we note 
that the peak is located at $p \approx gEt/2$. 
Thus the $p$ prefactor in \eqref{rhoexplicit} contributes one factor of $t$. In the 
integrand, the initial conditions for the variables
${\dot \phi}_{\bfp,\pm}$ and ${\dot \psi}_{\bfp,\pm}$ are proportional to
$p$ as in \eqref{dotic} and since the $p$ that contributes to the energy 
integral grows proportional to $t$, and the variables enter quadratically in 
the energy integral, the peak of the spectral energy density grows as $t^3$.

\begin{figure}
      \includegraphics[width=0.45\textwidth,angle=0]{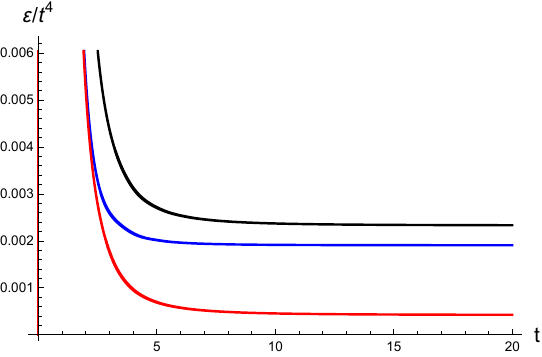}
 \caption{
Energy density rescaled by $t^4$, $:{\calE}:/(L^3 t^4)$, vs. time in the $\phi$ excitations 
(blue curve),
$\psi$ excitations (red curve), and in total (black curve). The flat curves at late
times confirm that $:{\cal E}: \propto t^4$ as also indicated in Fig.~\ref{rhovsp}.
}
\label{evstime}
\end{figure}

The $t^4$ growth of the total energy density is further confirmed in Fig.~\ref{evstime} 
where we plot the total energy density divided by $t^4$ vs. time.

\begin{figure}
      \includegraphics[width=0.45\textwidth,angle=0]{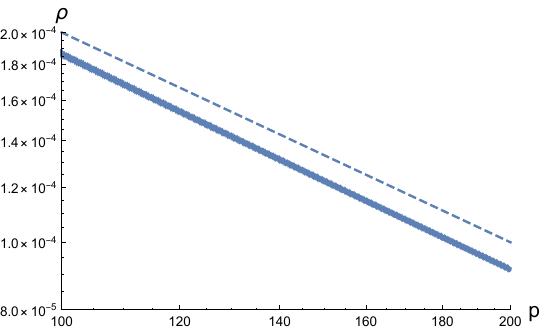}
 \caption{
Log-log plot showing the asymptotic behavior of $\rho(p,t)$ vs. $p$ for $t=2t_f/3$ with $t_f=20$. 
The dashed line shows a $1/p$ fall off.
}
\label{rhovspzoom}
\end{figure}

To obtain the total energy density, $:{\cal E}:$, we have integrated over
$p \in [0,p_c]$ where $p_c \gg gE t$ is a cutoff. To study the
dependence of our result on the cutoff, we zoom into the
large $p$ behavior of $\rho(p,t)$, shown in Fig.~\ref{rhovspzoom}.
This gives $\rho (p,t) \propto 1/p$ at large $p$ and the total energy diverges
logarithmically as the cutoff $p_c$ is taken to infinity. However this ultraviolet
divergence will be controlled once asymptotic freedom is taken into account.
To see this in more detail, note that $\rho(p,t)$ in Fig.~\ref{rhovsp} has a
dominant peak structure followed by the $1/p$ fall off,
which after integration over $p$,  lead to an asymptotic contribution to the energy 
density given
by $ g^4\,\log(p_c/M)$,  where $M$ is a renormalization scale.  
However the asymptotic value of the coupling constant at the cut-off scale evolves 
from its value $g_{M}$ at the renormalization scale as \cite{Peskin:1995ev},
 \be
 g^2={g_{M}^2\over 1+g_{M}^2 \log(p_c/M)}\,,
 \ee
implying,
\be
\lim_{p_c\to \infty} g^4\,\log(p_c/M) \to 0\,.
\ee
Hence the $1/p$ tail contribution to the integration vanishes in the $p_c \to \infty$ 
limit once we take the running of $g^2$, {\it i.e.} asymptotic freedom, into account.

\section{Number density production}
\label{numberdensity}

The number density of produced particles, $n$, is ambiguous at intermediate
times though there is no ambiguity in the infinite time limit if the electric field is switched
off~\cite{Dabrowski:2016tsx}. Here we 
adopt the definition that the number density is
given by the energy density for each mode divided by the frequency of that mode.
Then using \eqref{totalENO} we have,
 \ba
n &=& 8\pi \int_0^\infty dp\, p \int_0^1 du\,   
\sum_{s=\pm}
\biggl [ 
\frac{\left | {\dot \phi}_{\bfp,s}  - i \omega_{\phi,\bfp,s}  \, \phi_{\bfp,s} \right | ^2} {\omega_{\phi,\bfp,s}}  \nn \\
&& \hskip 2.5 cm
+ \frac{ \left | {\dot \psi}_{\bfp,s} -i \omega_{\psi,\bfp,s} \, \psi_{\bfp,s}\right  |^2}{\omega_{\psi,\bfp,s}}   \biggr ]
\label{totalnNO}
\ea
The expressions for the frequencies are given in \eqref{omphi} and \eqref{ompsi}. The issue
is that at any given time, there is a momentum mode for which the frequency vanishes. Thus
the integrand in \eqref{totalnNO} is singular at all times. 

To determine if the singularity is integrable, we focus our attention on the $\phi_{\bfp,-}$ sector
for which the number density is
\be
n_{\phi,-} = 8\pi \int_0^\infty dp\, p \int_0^1 du\,   
\frac{\left | {\dot \phi}_{\bfp,-}  - i |p-gEt u| \, \phi_{\bfp,-} \right | ^2} {|p-gEt u|}  
\label{nphi-}
\ee
First consider the numerator of the integrand. For $p=gEt u$ it is simply $|{\dot \phi}_{\bfp,-}|^2$.
Non-vanishing energy in the $\phi_{\bfp,-}$ excitations implies that $|{\dot \phi}_{\bfp,-}|^2 \ne 0$.
So the singularity structure of the integral is
\be
n_{\phi,-} = 8\pi \la p | {\dot \phi}_{\bfp,-} |^2 \ra \int_0^\infty dq \int_0^1 du\,   
\frac{1} {|q-u|}  
\label{nsing}
\ee
where $q \equiv p/gEt$ and $\la \cdot \ra$ denotes an effective value along the singular curve, 
$p=gEtu$ (or $q=u$), in the integration plane. By transforming
integration variables to $x_\pm =q \pm u$ it is clear that the integral is logarithmically
divergent due to the singularity along $x_-=0$.

We note that the divergence in particle number density arises at low energy where
the frequencies vanish. It is worth emphasizing that this is a particular behavior of 
the massless theory and will not be present in massive cases such as QED. 
In the full interacting theory we can expect this infrared divergence to be resolved due to
confinement effects. At such low frequencies,  the ``soft'' gluons are confined and are only present
in massive glueball states.  
In our analysis, as discussed in Sec.~\ref{setup}, we are examining where the road
goes when we ignore confinement effects. It is interesting that the calculation without
confinement leads to a divergent number density of excitations, suggesting a 
self-inconsistency.

\section{Adiabatic case}
\label{adiabatic}

Often in the Bogoliubov method, the background is taken to turn on adiabatically,
survive for a certain time period, and then slowly turn off. Then the asymptotic
vacua are unambiguously defined and the total energy density of particles produced
is evaluated at $t \rightarrow \infty$.

\begin{figure}
      \includegraphics[width=0.45\textwidth,angle=0]{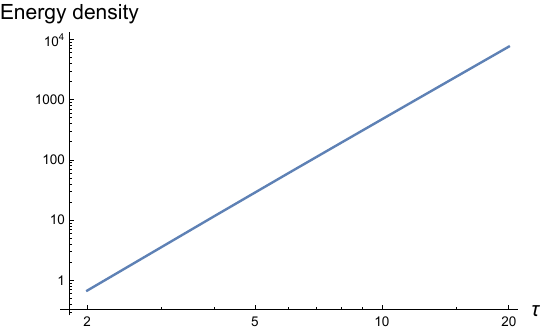}
 \caption{
Log-log plot of $:{\cal E}:$ vs. $\tau$ for the adiabatic case.
}
\label{evstau}
\end{figure}

We have also treated the adiabatic case. Now our choice for $A(t)$ is,
\be
A(t) = E \tau \left [ \tanh \left ( \frac{t-t_E}{\tau} \right ) - \tanh \left ( \frac{-t_E}{\tau} \right ) \right ] 
\ee
where $t_E$ is some large time at which the electric field is maximum and $\tau$ is the duration
for which the field is turned on. The electric field is given by,
\be
{\dot A} = E\, {\rm sech}^2 \left ( \frac{t-t_E}{\tau} \right ) 
\label{Epulse}
\ee
Our analysis from the previous sections remains unchanged except that the frequencies in
\eqref{omphiadiabatic} and \eqref{ompsiadiabatic} now become
\be
\omega_{\phi,\bfp,\pm} = +\sqrt{(p\pm gA(t) u )^2}
\label{omphiadiabatic}
\ee
\be
\omega_{\psi,\bfp,\pm} = +\sqrt{(p_z \pm gA(t))^2+p_\perp^2}
\label{ompsiadiabatic}
\ee
We have repeated the numerical analysis in this case and show the results for the
energy density versus $\tau$ in Fig.~\ref{evstau}. Once again we find 
$:{\cal E}: \propto \tau^4$.

We can also consider the number density of particles in the adiabatic case.
From Eq.~\eqref{omphiadiabatic} we see that the mode with $p=gA(\infty) u$
has vanishing frequency $\omega_{\phi,\bfp,-}$ as $t\to \infty$. Similarly the mode with 
$p_z=gA(\infty)$, $p_\perp=0$ has vanishing frequency $\omega_{\psi,\bfp,-}$ as $t\to \infty$.
Therefore the singularity discussed in Sec.~\ref{numberdensity} is present
even with the short duration electric field of \eqref{Epulse}.

\section{Conclusions}
\label{conclusions}

We have considered the fate of a uniform color electric field in 
a pure gauge Yang-Mills theory 
with SU(2) gauge group\footnote{The gauge group is not important for our analysis since
SU(N) models have SU(2) subgroups.}. Our approach assumes initial conditions with a
uniform electric field and quantum excitations in their non-interacting ground state. We
then truncate the excitations to quadratic order in the action and evolve the system using
the CQC.
 
We find that there is rapid particle (gluon) production for which we are able to characterize the 
energy spectrum and its time dependence as shown in Fig.~\ref{rhovsp}. Production occurs for 
modes in a range $p \in [0,gEt]$ with peak production at $p \approx gEt/2$. The amount of energy
produced in an interval $dp$ grows as $t^3$. We can understand this growth in terms of
phase space factors that give one power of $t$, and the square of the amplitude of vacuum 
fluctuations that are proportional to $p^2$, and hence grow as $t^2$. The $t^4$ dependence 
we find is in sharp contrast to usual Schwinger pair production, as in \eqref{schwinger}, for 
which the energy grows linearly with time. However, in contrast to the usual Schwinger
pair productions of massive particles, gauge excitations are massless and this may be 
sufficient to explain the different production rates.

To the order in which we perform our calculations, the coupling constant does not run
with energy scale. However, we encounter an ultraviolet divergence in our calculation 
of the energy produced in excitations. We have argued that the divergence would get 
controlled if we were to properly account for the running of the coupling constant at 
high energies (asymptotic freedom). In calculating the particle number density we also 
encounter an infrared divergence that we argue will be absent in the confining theory. 
In other words, if we ignore confinement, our analysis implies that the particle number 
density diverges, which we interpret as an indication of a lack of self-consistency
of the unconfined assumption.

Our analysis leaves open several directions of interest such as the backreaction of particle
production on the background color electric field. If we simply use energy conservation as 
a guide, the background electric field has an energy density proportional to $E^2$ whereas 
the energy density in excitations grows as $g^4E^4 t^4$. Equating these two gives
the decay time for the electric field to be $\tau \sim (g\sqrt{E})^{-1}$. 
However, we cannot exclude the possibility that the electric field will be antiscreened
as argued for non-Abelian gauge theories (see for example Sec.~16.7 of~\cite{Peskin:1995ev}),
in which case the electric field strength would actually increase with time.
We hope to investigate this issue by a more detailed analysis in the future.

\acknowledgements
We thank Jan Ambj\o rn, Mainak Mukhopadhyay, Subodh Patil, Doug Singleton and  George Zahariade
for feedback.
T.V.  was supported by the U.S. Department of Energy, Office of High Energy Physics, 
under Award No.~DE-SC0019470 and C.C.  by the National Science Foundation Award 
No.~PHY-2012195 at Arizona State University.

\appendix

\section{Classical-Quantum Correspondence}
\label{appA}

The CQC is equivalent to the Bogoliubov transformation method for calculating particle
production in the case of a time dependent but spatially homogeneous background. Once
the model is discretized, each mode is equivalent to a simple harmonic oscillator as
in Sec.~\ref{discretization}. So we only need to demonstrate the method for a simple
harmonic oscillator. 

Consider the Hamiltonian for a simple harmonic oscillator of mass $m$ -- which can also
be time dependent in the general case -- with an arbitrary time dependent frequency $\omega (t)$,
\be
H = \frac{p^2}{2m} + \frac{m}{2} \omega^2(t) x^2
\ee
Then the quantum operators $x$ and $p$ can be written as
\be
x = z^* a_0 + z a_0^\dag, \ \
p = m( {\dot z}^* a_0 + {\dot z} a_0^\dag )
\ee
where $a_0$ and $a_0^\dag$ are the initial annihilation and creation operators defined
in terms of the initial position operator $x_0$ and momentum operator $p_0$,
\be
a_0 = \frac{p_0-im\omega_0 x_0}{\sqrt{2m\omega_0}}, \ \ 
a_0^\dag = \frac{p_0+im\omega_0 x_0}{\sqrt{2m\omega_0}}
\ee
where $\omega_0$ is the frequency at the initial time, and $z(t)$ is a complex-valued 
c-number function of time. Using the Heisenberg equations for $x$ and $p$ we find
that $z(t)$ must satisfy
\be
{\ddot z} + \omega^2 z = 0
\ee
The initial conditions follow from the above relations and are,
\be
z_0 = \frac{-i}{\sqrt{2m\omega_0}}, \ \ {\dot z}_0 = \sqrt{\frac{\omega_0}{2m}}
\ee
The Wronskian is constant and fixed by the initial conditions,
\be
m( z^* {\dot z} - z {\dot z}^* ) = i
\ee

Other quantum operators can be re-written in terms of $z(t)$, $x_0$ and $p_0$. 
Expectation values are written entirely in terms of $z(t)$. In particular,
the energy is
\be
\la H \ra = \frac{1}{2} m |{\dot z}|^2 + \frac{1}{2} m \omega^2 |z|^2
\ee
This can also be written as
\be
\la H \ra = \frac{m}{2} | {\dot z} - i \omega z |^2 - \frac{i}{2} m \omega( z^* {\dot z} - z {\dot z}^* )
\ee
The second term is the Wronskian and hence is a constant determined by the initial
conditions. In the quantum simple harmonic oscillator this term corresponds to the
ground state energy $\omega /2$. So the energy in the excitations is given by
\be
:{\cal E}: = \frac{m}{2} | {\dot z} - i \omega z |^2 
\ee
which is what we use in \eqref{calE1NO} and \eqref{calE2NO}.

\newpage

\bibstyle{aps}
\bibliography{paper}

\end{document}